\documentclass[%
 reprint,
 amsmath,amssymb,
 aps,%pra,
]{revtex4-1}

\usepackage{graphicx}% Include figure files
\usepackage{dcolumn}% Align table columns on decimal point
\usepackage{bm}% bold math
\begin{document}

\preprint{APS/123-QED}

\title{ Results in the Spontaneous Annihilation of the Cosmological Constant}% Force line breaks with \\
%\thanks{A footnote to the article title}

\author{DongBong Yang$^*$ and Heon-Ju Lee$^\dagger$}
\affiliation{%
$^*$Department of Science, World Culture Open,\\
 260-88 Seongbuk-dong, Seongbuk-gu, Seoul, Korea\\
$^\dagger$Department of Nuclear and Energy Engineering, Jeju National University,\\
102 Jejudaehak-ro, Jeju-si, Jeju 63243, Korea\\
dongbongyang@hanmail.net korea\\
}%


\date{\today}% It is always \today, today,

\begin{abstract}
We suggest a new formula, which allows the Schwarzschild's solution and the Einstein radius to be applied to the dynamic universe, when our universe is hypothetically regarded as a single dynamic black hole. In this study, a cosmological constant problem is solved in the simplest manner, while we find excellent agreements with observation. We adopt a model, wherein k=0, $\Lambda$$\ne$0, and $\Omega$=1 to interlock $\Lambda$ with critical density of the black hole of our universe $\rho_{\rm{cBH}}$, thereby presenting complimentary relation between $\Lambda$ and Bekenstein-Hawking entropy S$_{\rm{BH}}$.

\begin{description}
\item[PACS numbers]
98.80.k, 98.80.Bp, 98.80.Es
\item[Structure]
$\Omega$, Cosmological constant, Black hole entropy
\end{description}
\end{abstract}

\pacs{Valid PACS appear here}% PACS, the Physics and Astronomy

%Use showkeys class option if keyword

\maketitle

%\tableofcontents

\section{\label{sec:level1}Background}
Unexpected results are discovered over a process to analyze and examine a hierarchical problem of elementary particles and their free parameters. We generate a large number of phenomenological models by applying an abduction method to a data set of Particle Data Group (PDG). Through the models, we acquire certain relation among three key parameters: the first key parameter is mass of three families of quarks and neutrinos from fundamental particles; the second key parameter is three components of the universe, $\Omega_{tot}$=1, i.e. (0.683+0.268+0.049) $\rho_{c}$; and the third key parameter is the Leech lattice (196560 = 98304+97152+1104), which has three geometric components related to the Monster symmetry. The maximum radius of a black hole of our universe R$_{\rm{smax}}$ which is extrapolated from (a) a certain combination of quarks and neutrinos, based on the three key parameters, and (b) relation of CMB, provides marvelous information with regard to an initial condition of our universe and the cosmological constant problem\cite{r1}. In this study, all R$_{\rm{smax}}$ translate R$_{\rm{s}}$, which has a static or isolated universal property, in a new way with regard to the dynamic universe, so that we regard R$_{\rm{smax}}$ as R$_{\rm{s}}$, which refers to the maximum critical radius, in an observable range.

Such new information is a result, which is generated by a predictive analysis model through machine learning and a self-developed artificial intelligence mechanism in order to analyze Planck collaboration cosmological parameters in a sophisticated manner. In this study, we set two fixed parameters: the Newtonian gravitational constant G as a component of the stress-energy tensor and the maximum radius of the black hole of our universe $\rm R_{smax}$.

To begin with, we introduce a so-called ``Zero Zone unit system postulate \cite{r2}'', having invariance dimensionality symmetrical to unit transformation, while maintaining uniformity of all SI unit systems. The postulate realizes the idea, suggested by the Buckingham $\pi$ theorem\cite{r3}, which presents that all natural phenomena and physical laws can be expressed as pure numbers. We apply Eq.(\ref{eq:11}) to all equations in this study and simplify them.
%{1-1}
\begin{equation}\label{eq:11}\tag{1.1}
[c]=[\mathit{h}]=[1\mathrm{s}]=[e/\mathit{m}_{\mathrm{e}}]=[\textit{k}_{B}]=[N_{\mathrm{A}}]=[b]=1
\end{equation}

Implications of the postulate can be stated as a ratio of dimensionless numbers. Namely, formulae can be regarded as physically identical when their calculated values are the same, i.e. 1 = 1s = 1Hz and 2 = 2s = 2Hz$\dots$, albeit they have different structures. In cosmology, a dimensionless number `1' is interpreted as a state, wherein the early universe had high order, at one second after the inflation.

The first fixed parameter, the Newtonian gravitational constant G, is presented as follows.

%(1-2)
\begin{equation}\label{eq:12}\tag{1.2}
G\approx6.673~384~392\dots\times10^{-11} ~~ {\mathrm{m^3 \cdot kg^{-1} \cdot s^{-2}}}
\end{equation}

%Eq. (1-2-1)
\begin{equation}\label{eq:121}\tag{1.2.1}
\begin{aligned}
&\rm{Planck~mass~{\mathit{M_{\rm P}}}}\\
&(\hbar c/G)^{1/2}=(\frac{1}{2\pi G})^{1/2} \approx 2.176~508~699\dots\times10^{-8} ~~ {\mathrm{kg}}
\end{aligned}
\end{equation}

%Eq. (1-2-2)
\begin{equation}\label{eq:122}\tag{1.2.2}
\begin{aligned}
&\rm{Planck~length~{\mathit{L_{\rm P}}}}\\
&(\frac{\hbar G}{c^{3}})^{1/2}=(\frac{G}{2\pi})^{1/2} \approx 1.616~199~666\dots\times 10^{-35}~~{\mathrm{m}}
\end{aligned}
\end{equation}

%Eq. (1-2-3)
\begin{equation}\label{eq:123}\tag{1.2.3}
\begin{aligned}
&\rm{Planck~energy~density~{\mathit{\rho_{\rm P}}}}\\
&(\frac{E_{\rm P}}{L^{3}_{\rm P}})=(\frac{M_{\rm P}}{L^{3}_{\rm P}}) \approx 5.155~554~340\dots\times 10^{96}~~{\mathrm{kg \cdot m^{-3}}}
\end{aligned}
\end{equation}

%Eq. (1-2-4)
\begin{equation}\label{eq:124}\tag{1.2.4}
G=(\frac{L_{\rm P}}{M_{\rm P}}),~2\pi L^{2}_{\rm P}=G,~(\frac{G}{L_{\rm P}})^{2}=2\pi G
\end{equation}

\section{\label{sec:level2}Rescaled Friedmann equation}
Friedmann considered the subsequent Einstein field equation, which includes the cosmological constant as a starting point\cite{r4}.

%Eq. (2-1)
\begin{equation}\label{eq:21}\tag{2.1}
G_{\mu\nu}+\Lambda g_{\mu\nu}=\frac{8\pi G}{c^4}T_{\mu\nu}
\end{equation}

Eq.(\ref{eq:21}) is simplified by

%Eq. (2-2)
\begin{equation}\label{eq:22}\tag{2.2}
H^{2}=(\frac{\dot{a}}{a})^{2}=\frac{8\pi G}{3}\rho-\frac{kc^{2}}{a^{2}}+\Lambda\frac{c^{3}}{3}
\end{equation}

For further simplification of Eq.($\ref{eq:22}$), Friedmann formulated the following formulae, under the assumption that k=0 and $\Lambda$=0.

%Eq. (2-3)
\begin{equation}\label{eq:23}\tag{2.3}
H^{2}=\frac{8\pi G}{3}\rho
\end{equation}
Or,
%Eq. (2-4)
\begin{equation}\label{eq:24}\tag{2.4}
\rho=\frac{3H^{2}}{8\pi G}
\end{equation}

In the Einstein-de Sitter Model, wherein E = 0, the flat universe, k = 0, corresponds to the Friedmann model. In this case $\rho$$\rightarrow$$\rho_{c}$, so that an independent hypothesis is required.

\section{\label{sec:level3}Hypothesis and Spontaneous Annihilation Mechanism}
In the actual observation of the universe \cite{r5},\cite{r6}, the cosmological constant, $\Lambda$, is not equal to zero, $\Lambda$$\ne$0. Therefore, in this study, we have two hypotheses to include $\Lambda$ and satisfy the observation data and the Einstein field equation. The first one is that k = 0, $\Lambda$$\ne$0, and $\Omega$ = 1, since we adopt the theories \cite{r7},\cite{r8}, which provide meaningful statements with regard to the rescaled Friedmann model and the density parameter, $\Omega$= 1. We come up with the second hypothesis in order to precisely match relation among actual density $\rho$, critical density $\rho_{c}$ , and critical density of the black hole of our universe $\rho_{cBH}$, each of which is a cosmological parameter. That is,

%Eq. (3-1)
\begin{equation}\label{eq:31}\tag{3.1}
\Omega=\frac{\rho}{\rho_{c}}=\frac{8\pi G}{3H^{2}}\rho=1
\end{equation}

%Eq. (3-2)
\begin{equation}\label{eq:32}\tag{3.2}
\rho=\rho_{c}=\rho_{cBH}
\end{equation}

In the Hubble's law, which is expressed by v=H$\cdot$r, when velocity of a galaxy is v$\rightarrow$c, in a cosmological horizon (particle horizon), the following transformation is possible.

%Eq. (3-3)
\begin{equation}\label{eq:33}\tag{3.3}
r=\frac{c}{H}=\frac{1}{H}
\end{equation}

According to the hypothesis of Eq.(\ref{eq:32}), when we assume that our universe is the single dynamic black hole in Eq.(\ref{eq:33}), then moving time r in the Hubble time, t$\rm_{H}$, is closer to $\bf R_{s}$, having the maximum critical radius $\bf R_{smax}$, namely, r$\rightarrow$$\bf R_{s}$. Therefore, we have consistent relation among $\bf R_{s}$, $\rho_{\rm{cBH}}$, $\rho_{c}$ and $\rho$  in the hypothesized dynamic black hole of our universe. The FLRW cosmological model describes a boundary condition\cite{r9} with regard to a solution of the particle horizon $\textit H_{\rm P}$. The particle horizon is similar to an event horizon, but has little difference. It is noteworthy that in this study, the Schwarzschild radius, R$_{s}$, which refers to the event horizon is different from R$_{s}$, which refers to the particle horizon. We will describe the difference, corresponding to an upper bound concept \cite{r10} of entropy in the black hole. All symbols in bold font are cosmological parameters related to the $\textit{comoving universe}$, e.g. $\bf M$, and $\bf R_{s}$. Given aforementioned statements, Eq.(\ref{eq:31}) can be re-described by

%Eq. (3-4)
\begin{equation}\label{eq:34}\tag{3.4}
\Omega=\frac{\rho}{\rho_{c}}=\frac{8\pi G}{3H^{2}}\rho=\frac{8\pi G{\mathbf{R_{s}}}^{2}}{3}\rho_{c}=1
\end{equation}

In addition, Eq.(\ref{eq:33}) clarifies the meaning of Eq.(\ref{eq:24}), and relation between $\bf R_{s}$ and $\bf M$ is proportional. Therefore, $\rho$ is transformed as below.

%Eq. (3-5)
\begin{equation}\label{eq:35}\tag{3.5}
\begin{aligned}
&\rho=\frac{3H^{2}}{8\pi G}=\frac{3}{8\pi G{\mathbf{R_{s}}}^{2}}=\frac{3}{32\pi G^{3}{\mathbf{M}}^{2}}=\rho_{c}=\rho_{cBH}\\
&\because{\mathbf{R_{s}}}=2G{\mathbf{M}}/c^{2}=2G{\mathbf{M}}
\end{aligned}
\end{equation}

In Eq.(\ref{eq:35}), it is notable that
%Eq. (extra)
\begin{displaymath}
\rho = \rho_{c} = \rho_{cBH}\varpropto\frac{1}{{\mathbf{R_{s}}}^{2}}, \varpropto\frac{1}{{\mathbf{M}}^{2}}
\end{displaymath}

After completing his field equation, Einstein acquired information of the Hubble's expanding universe, and corrected his static universe concept in the following way. Namely, it is presented that the cosmological constant, $\Lambda$ , is proportional to the following terms.

%Eq. (3-6)
\begin{equation}\label{eq:36}\tag{3.6}
\Lambda \varpropto (\frac{1}{R^{2}}) \varpropto (\frac{\kappa\rho}{2})
\end{equation}

where

$\kappa$=Einstein's constant

Given Eq.(\ref{eq:35}),
\begin{displaymath}
(\frac{1}{R^{2}}) \rightarrow(\frac{1}{{\mathbf{R_{s}}}^{2}})
\end{displaymath}
in Eq.(\ref{eq:36}), while it seems that
\begin{displaymath}
\rho\rightarrow\rho_{c}, \rho_{cBH}
\end{displaymath}

 If the implication of Eq.(\ref{eq:36}) is correct, it is expected that $\kappa$=2. This case seems to be associated with the $\textit H_{\rm P}$ boundary condition.

Generally, Planck energy density $\rho_{\rm P}$
\begin{displaymath}
\rho_{\rm P}=(\frac{E_{\rm P}}{L^{3}_{\rm P}})=(\frac{M_{\rm P}}{L^{3}_{\rm P}})
\end{displaymath}
is used in a unit of the cosmological constant. Considering observation data and Eqs.(\ref{eq:35}),(\ref{eq:36}) simultaneously $\Lambda$  can be expressed as $\rho_{c}$ or $\rho$  with regard to Planck energy density $\rho_{\rm P}$. In this case, if an annihilation factor is set as z$^{*}$ , then z$^{*}$ is expressed as

%Eq. (3-7)
\begin{equation}\label{eq:37}\tag{3.7}
z^{*}\Lambda = (\frac{z^{*}\rho}{M_{\rm P}/L^{3}_{\rm P}}) = (\frac{z^{*}\rho_{c}}{M_{\rm P}/L^{3}_{\rm P}})
\end{equation}

In Eq.(\ref{eq:37}),
\begin{displaymath}
\rho=\frac{3H^{2}}{8\pi G}=\frac{3}{8\pi G{\mathbf{R_{s}}}^{2}}=\frac{3}{32\pi G^{3}{\mathbf{M}}^{2}}=\rho_{c}=\rho_{cBH}
\end{displaymath}
Therefore, $z^{*}\Lambda$ can be rewritten as follows.

%Eq. (3-8)
\begin{equation}\label{eq:38}\tag{3.8}
z^{*}\Lambda=\frac{z^{*}3L_{\rm P}^{3}}{8\pi G{\mathbf{R_{s}}}^{2}M_{\rm P}}=(\frac{z^{*}3L_{\rm P}^{3}}{32\pi G^{3}{\mathbf{M}}^{2}M_{\rm P}})
\end{equation}\\

The second fixed parameter, the black hole radius $\bf R_{s}$ of our universe, is shown in the following numerical value.
%Eq. (3-9)
\begin{equation}\label{eq:39}\tag{3.9}
{\mathbf{R_{s}}}\approx 1.721~944~132\dots\times 10^{26}~~ {\mathrm{m}}
\end{equation}

The numerical value of R$_{\rm s}$ is of the order of 10$^{10}$ light years

By using the numerical value.(\ref{eq:39}), we can obtain the critical density of the black hole of our universe $\rho_{cBH}$ as follows.

%Eq. (3-10)
\begin{equation}\label{eq:310}\tag{3.10}
\begin{aligned}
&\rho_{cBH}=\frac{3H^{2}}{8\pi G}=\frac{3}{8\pi G{\mathbf{R_{s}}}^{2}}=\frac{3}{32\pi G^{3}{\mathbf{M}}^{2}}\\
&\approx 5.421~365~353\dots\times 10^{-27}~~{\mathrm{kg \cdot m^{-3}}}
\end{aligned}
\end{equation}\\

Given the value from Eq.(\ref{eq:310}), it is expected that the number of hydrogen atoms is around 3.241 238 054 $\cdots$per m$^{3}$.

As such, mass of our universe $\bf{M}$ is multiplication of volume from the numerical value(\ref{eq:39}) and density from Eq.(\ref{eq:310}).

%Eq. (3-11)
\begin{equation}\label{eq:311}\tag{3.11}
\begin{aligned}
&{\mathbf{M}}=\frac{4\pi{\mathbf{R_{s}}}^{3}}{3}\times{\rho_{cBH}}\\
&\approx 1.159~456~489\cdots\times 10^{53}~~ {\mathrm{kg}}
\end{aligned}
\end{equation}

With regard to $\bf M$ in Eq.(\ref{eq:311}), if proton\\
m$_{\rm P}$ $\approx$ 1.672 621 776$\cdots\cdots$10$^{-27}$kg, then, the number of protons is around 6.931 970 548$\cdots\cdots$$\times$10$^{79}$ in $\bf{M}$, and this shows that observation data and theoretical prediction are approximately identical.

As such, a calculated dimensionless value of $\Lambda$  excepting for z$^{*}$ is defined by

%Eq. (3-12)
\begin{equation}\label{eq:312}\tag{3.12}
\Lambda \approx 1.051~558~182\dots\times 10^{-123}
\end{equation}\\

When we transform the processes of $\textit M_{\rm P}$$\rightarrow$$\textbf{M}$ and $\textit L_{\rm P}$$\rightarrow$ $\bf{R_{s}}$ (max.) into calculated values, then it is possible to obtain the following relation.

%Eq. (3-13)
\begin{equation}\label{eq:313}\tag{3.13}
\frac{\mathbf{M}}{M_{\rm P}}=\frac{\mathbf{R_{s}}}{2L_{\rm P}}\approx 5.237~139~237\cdots\times 10^{60}
\end{equation}

Eq.(\ref{eq:313}) implies that our universe was inflated more than 10$^{60}$ times at one second after the inflation, based on the Plank scale. It seems that the early universe maintains such order with scale invariance.

 $\Lambda$ seems to be proportional to the critical density of the black hole $\rho_{c}$ in Eqs.(\ref{eq:37}),(\ref{eq:38}). However, it is notable that, when density of black hole critical density $\rho_{c}$ reaches a lower bound along with flow of time, $\Lambda$ is inversely proportional to the maximum critical density radius $\bf R_{smax}$ and the maximum mass $\bf M_{max}$, thereby being minimized. This is because the black hole is the system, which has the maximum entropy of our universe $\bf S_{max}$\cite{r11} under the complementary relation, and it is acknowledged that the black hole has information analogous to that of thermodynamics \cite{r12}. Especially, with regard to the holographic principle, suggested by t'Hooft, Susskind \cite{r13} presented that all entities in a certain area of a space can be described by information pieces, distributed on a boundary surface. Briefly, Susskind and others suggested that a positive cosmological constant has surprising consequences, such as the finite maximum entropy of the observable universe. Consequently, in Eq.(\ref{eq:37}), the annihilation factor, z$^{*}$, is regarded as the most important parameter, which mediates between gravitational force and quantum mechanics. The annihilation factor, indeed, is the final result of a large number of phenomenological models, and it is confirmed that z$^{*}$ can be replaced by the Bekenstein-Hawking entropy formula $\rm S_{BH}$ in a precise manner. That is,

%Eq. (3-14)
\begin{equation}\label{eq:314}\tag{3.14}
z^{*}={\rm S_{BH}}=\frac{k_{\textrm{B}}c^{3}{\mathbf{A}}}{4\hbar G}
\end{equation}

In Eq.(\ref{eq:314}), $\bf A$ is an area of the event horizon of our universe, and $\bf{A= 4\pi R_{S}^{2}}$. Here, we regard $\bf A$ as a cosmological horizon, instead of the event horizon. When [$\textit{c}$]=[$\textit{h}$]=[$\textit k_{\textrm{B}}$]= 1 is applied, ${\rm S_{BH}}$ can be briefly expressed as follows.

%Eq. (3-15)
\begin{equation}\label{eq:315}\tag{3.15}
{\rm S_{BH}}=\frac{\pi \bf A}{2G}=\frac{\pi(4\pi{\mathbf{R_{s}}}^{2})}{2G}=\frac{4\pi^{2}(2G{\mathbf{M}})^{2}}{2G}=8\pi^{2}G{\mathbf{M}}^{2}
\end{equation}

%Eq. (3-16)
\begin{equation}\label{eq:316}\tag{3.16}
{\rm S_{BH}}=\frac{\bf A}{4L^{2}_{\rm P}}=\frac{4\pi G^{2}{\mathbf{M}}^{2}}{L^{2}_{\rm P}}=8\pi^{2}G{\mathbf{M}}^{2} (\because 2\pi L_{\rm P}^{2}=G)
\end{equation}

It is notable that, in comparison to Eq.(\ref{eq:38}), inverse relation is true: $\rm S_{BH}$$\varpropto$$\bf R_{s}^{2}$$\varpropto$$\bf M^{2}$   ,    .

In Eqs.(\ref{eq:315}),(\ref{eq:316}), it is possible to identify values of G, $\bf M$, and L$_{\rm P}$, so that a calculated dimensionless value of the observable maximum entropy of the universe $\rm S_{BH}$ is expressed in the following formula.

Roger Penrose used a symbol, `$\rm S_{\Lambda}$', which refers to the ultimate entropy and calculated its value. Our calculated value is nearly identical to the Penrose's result. Roger Penrose wrote, ``With the observed value of $\Lambda$, the temperature $T_{\Lambda}$ would have the absurdly tiny value ~10$^{-30}\rm K$, and the entropy S$_{\Lambda}$ would have the huge value ~$\sim 3\times 10^{122}$\cite{r14}.''

%Eq. (3-16-1)
\begin{equation}\label{eq:3161}\tag{3.16.1}
{\rm S_{BH}}=3.566~136~483\dots\dots\times 10^{122}
\end{equation}

Interaction between $\Lambda$ and S$_{\rm BH}$ from Eq.(\ref{eq:316}) is completely annihilated in a spontaneous manner. That is,

%Eq. (3-17)
\begin{equation}\label{eq:317}\tag{3.17}
\Lambda\cdot{\rm S_{BH}}=\frac{3L^{3}_{\rm P}}{32\pi G^{3}{\mathbf{M}}^{2}M_{\rm P}}\cdot\frac{4\pi G^{2}{\mathbf{M}}^{2}}{L_{\rm P}^{2}}=\frac{3L_{\rm P}}{8G\cdot M_{\rm P}}=\frac{3}{8}
\end{equation}
or
%Eq. (3-17-1)
\begin{equation}\label{eq:3171}\tag{3.17.1}
\frac{\rho_{c}}{M_{\rm P}/L_{\rm P}^{3}}\cdot\frac{4\pi G^{2}{\bf{M}}^{2}}{L_{\rm P}^{2}}-\frac{3}{8}=0
\end{equation}

\section{\label{sec:level4}Discussion and Conclusion}
Did any special initial state exist in our universe? In the Eq.(\ref{eq:3171}), if E = 0, and curvature	k=0, then $\rho$$\rightarrow$$\rho_{c}$. Namely, if we re-write the irreducible fraction, 3/8, as
\begin{displaymath}
\rho=\rho_{c}=\rho_{cBH}=\frac{3}{32\pi G^{3}{\mathbf{M}}^{2}}
\end{displaymath}
then, Eq.(\ref{eq:317}) regresses to the original Eq.(\ref{eq:35}). It implies that the maximum entropy of our universe S$_{max}$ encounters $\Lambda$ expressed by $\rho_{cBH}$, then this becomes an unobservable process. Namely, when entropy spontaneously collapses to satisfy Eqs.(\ref{eq:31}),(\ref{eq:32}) simultaneously, then this means that the special initial state exists with high order.

It shows agreements with the observation data and seems to contribute to find valuable information with regard to a question why entropy of our universe is low \cite{r15}. Namely, when the spontaneous annihilation mechanism of the cosmological constant is excluded, it is possible to come up with a question whether entropy can decrease, as the universe re-collapses to explain the low entropy of the early universe \cite{r16}.

Followed by Eq.(\ref{eq:317}), we present relation between entropy and the temperature. That is, when a formula of the Bekenstein-Hawking entropy $\rm S_{BH}$ is combined with a formula of the Hawking radiation temperature $\rm T_{BH}$, the result is shown below.

%Eq. (4-1)
\begin{equation}\label{eq:41}\tag{4.1}
\begin{aligned}
&{\rm S_{BH}}\cdot{\rm T_{BH}}\\
&=\frac{k_{\mathrm{B}}c^{3}{\mathbf{A}}}{4\hbar G}\cdot\frac{\hbar c^{3}}{8\pi GMk_{\mathrm{B}}}=8\pi^{2}GM^{2}\cdot\frac{1}{16\pi^{2}GM}=\frac{1}{2}M
\end{aligned}
\end{equation}

Here, we can have a calculated value of ${\rm T_{BH}}$ by substituting $\textit M$ with $\bf M$ in $\rm T_{BH}$=$\frac{1}{16\pi^{2}GM}$. When the calculated value is transformed into thermodynamic temperature, K, then it is possible to calculate the lower bound temperature. Roger Penrose used the symbol, `$T_{\Lambda}$' for his calculations, and our calculation is nearly equal to Penrose's numerical factor \cite{r14}.

%Eq. (4-1-1)
\begin{equation}\label{eq:411}\tag{4.1.1}
\begin{aligned}
{\rm T_{BH}}&=\frac{\hbar c^{3}}{8\pi G{\mathbf{M}}k_{\mathrm{B}}}=\frac{1}{16\pi^{2}G{\mathbf{M}}}\\
&\approx 1.058~241~232\dots\dots\times 10^{-30}~~{\mathrm{K}}
\end{aligned}
\end{equation}

Given Eqs.(\ref{eq:317}),(\ref{eq:41}), the final implication is that when entropy shows a waveform graph along with the flow of time, energy distribution, during the contraction/expansion phases, precisely becomes half, thereby generating a symmetrical pattern. However, during the contraction phase, energy is absorbed into an empty space and disappears. Therefore, only the expansion phase entropy can be observed, and thus it seems to show an asymmetrical structure in the direction of the arrow of time.

Eq.(\ref{eq:317}) implies the annihilation mechanism of $\Lambda$, which is generated by quantum perturbation in a vacuum, and coexistence of the initial and the final conditions of our universe. This means that we expand the theory: the critical density and the actual density are precisely matched so as to be autonomously coordinated with one another\cite{r17}. Namely, the upper bound of the black hold radius R$_{\rm smax}$, which corresponds to the maximum entropy S$_{\rm max}$, or its black hole volume is close to the lower bound, this means that a concept of the Bing Bang can be potentially replaced by Bouncing cosmology \cite{r18}. This idea provides a philosophical issue by referring to related articles.
The final relation among $\rm S_{BH}$, ${\rm T_{BH}}$, $\Lambda$ , G, and $\bf R_{s}$ is presented by

%Eq. (4-2)
\begin{equation}\label{eq:42}\tag{4.2}
\frac{\rm S^{2}_{BH}\cdot{\rm T_{BH}}\cdot \Lambda\cdot G}{{\mathbf{R_{S}}}}=\frac{3}{32}
\end{equation}

We verify Eq.(\ref{eq:42}) in the following way. If we express the formula in relation to $\Lambda$, then the result is presented as below.

%Eq. (4-2-1)
\begin{equation}\label{eq:421}\tag{4.2.1}
\Lambda=\frac{3{\mathbf{R_{s}}}}{32\cdot{\rm S^{2}_{BH}}\cdot {\rm T_{BH}}\cdot G}=\frac{6G{\mathbf{M}}\cdot 16\pi^{2}G{\mathbf{M}}}{32\cdot 64\pi^{4}G^{3}{\mathbf{M}}^{4}}=\frac{3}{64\pi^{2}G{\mathbf{M}}^{2}}
\end{equation}

In the Eq.($\ref{eq:36}$), when we exclude the annihilation factor, z$^{*}$, and describe $\Lambda$ only, then
\begin{displaymath}
\Lambda=(\frac{\rho}{M_{\rm P}/L^{3}_{\rm P}})=(\frac{\rho_{c}}{M_{\rm P}/L^{3}_{\rm P}})
\end{displaymath}
In addition,
\begin{displaymath}
\rho=\rho_{c}=\rho_{cBH}=\frac{3}{32\pi G^{3}{\mathbf{M}}^{2}}
\end{displaymath}
Therefore, Eq.(\ref{eq:421}) can be described as follows.

%Eq. (4-2-2)
\begin{equation}\label{eq:422}\tag{4.2.2}
\frac{3}{64\pi^{2}G{\mathbf{M}}^{2}}=\frac{3L_{\rm P}^{3}}{32\pi G^{3}{\mathbf{M}}^{2}{M_{\rm P}}}
\end{equation}

Eq.(\ref{eq:422}) can be simply stated by
\begin{displaymath}
\frac{1}{2\pi}=\frac{L_{\rm P}^{3}}{G^{2}M_{\rm P}}=\frac{L_{\rm P}^{2}}{G}\\
(\because G\cdot M_{\rm P}=L_{\rm P})
\end{displaymath}

The Einstein's static universe shows the following equation in the Friedmann equation with regard to the Einstein radius E$_{R}$.

%Eq. (4-3)
\begin{equation}\label{eq:43}\tag{4.3}
E_{R}=\frac{c}{\sqrt{4\pi G\rho}}
\end{equation}

When we apply c = 1, relation between the dynamic universe $\bf R_{S}$ and the static universe of the Einstein radius E$_{R}$ is expressed as below.

%Eq. (4-4)
\begin{equation}\label{eq:44}\tag{4.4}
\frac{{\mathbf{R_{s}}}}{E_{R}}=(\frac{3}{2})^{1/2}
\end{equation}

Eq.(\ref{eq:44}) is proven by the following procedure. As $\rho$ = $\rho_{c}$ = $\rho_{cBH}$, when both sides of Eq.(\ref{eq:44}) are squared, the result is as follows.

\begin{displaymath}
(\frac{{\mathbf{R_{s}}}}{E_{R}})^{2}= (2G{\mathbf{M}})^{2}\times 4\pi G\times\frac{3}{32\pi G^{3}{\mathbf{M}}^{2}}=\frac{3}{2}(QED)
\end{displaymath}

As a final result, extensively brief relation is produced between Eq.(\ref{eq:42}) and Eq.(\ref{eq:44}).

%Eq. (4-5)
\begin{equation}\label{eq:45}\tag{4.5}
\frac{3}{32}\cdot(\frac{{\mathbf{R_{s}}}}{E_{R}})^{2}=\frac{3}{32}\cdot\frac{3}{2}=\frac{9}{64}=(\frac{3}{8})^{2}=({\rm S_{BH}}\cdot\Lambda)^{2}
\end{equation}\\

%Conclusion
\section*{\label{sec:level5}Conclusion}
In this study, we find excellent agreements between (a) the annihilation mechanism of the cosmological constant $\Lambda$ and (b) a spontaneous decay process of the black hole maximum entropy S$_{\rm BH}$ itself in the consistent self-coordinating system, which makes $\rho$, $\rho_{c}$ and $\rho_{cBH}$  exactly identical at one second after the inflation of our universe. The procedure is a well-coordinated hidden mechanism, and it is unobservable. Therefore, the universe evolves from low entropy of our universe with regard to observation, so that the universe can be observed in the direction of the arrow of time. Despite this fact, energy and entire information are preserved.

\begin{acknowledgments}
 Authors are grateful to a Buckingham Machine research team for their contribution to establish an artificial intelligence system over a long haul. We also warmheartedly thank Dr. Dae Soo Han, a theoretical physicist who served in NASA for a long time, as his advice contributes to our theoretical foundation. Above all, we would like to express my deepest gratitude to Mr. Seok Hyun Hong, the former chairman of Joongangilbo for his financial support to build our research team.
\end{acknowledgments}



\end{document}